\documentclass[preprint,aps,floats,showpacs]{revtex4}

\usepackage{graphicx}
 \usepackage{epsfig}
\usepackage{pstricks}
\usepackage{pst-coil}
\def\PRL#1{{ Phys.\ Rev.\ Lett.} {\bf #1}}
\def\PRD#1{{ Phys.\ Rev.} {\bf D#1}}
\def\NPB#1{{ Nucl.\ Phys.} {\bf B#1}}
\def\PLB#1{{Phys.\ Lett.} {\bf B#1}}

\def\be{\begin{equation}}
\def\ee{\end{equation}}
\def\bea{\begin{eqnarray}}
\def\eea{\end{eqnarray}}

\def\sss{\scriptscriptstyle}
\def\Ls{{\sss L}}

\def\Rs{{\sss R}}
\def\Ns{{\sss N}}

\def\dl{\Delta_\Ls}

\def\dr{\Delta_\Rs}
\def\p{\phi}

{\newcommand{\lsim}{\mbox{\raisebox{-.6ex}{~$\stackrel{<}{\sim}$~}}}
{

\begin{document}

\preprint{IITB-TPH-0301}
\title{Leptogenesis with Left-Right domain walls}

\author{$\rm U ~A ~YAJNIK^{1}\footnote{Presented by UAY at 
PASCOS 2003, TIFR, Mumbai}, ~J ~CLINE^{2}, ~M ~RABIKUMAR^{1}$}

\affiliation{1 Indian Institute of Technology Bombay, Mumbai
400\thinspace076, India \\
2 McGill University, Montr\'eal, Qu\'ebec H3A 2T8, Canada\\
}

\date{March 30, 2003}

\begin{abstract}
The presence of domain walls separating regions of unbroken
$SU(2)_L$ and $SU(2)_R$ is shown to provide necessary conditions
for leptogenesis which converts later to the observed Baryon 
aymmetry. The strength of lepton number violation is related to
the majorana neutrino mass and hence related to current bounds
on light neutrino masses. Thus the observed neutrino masses 
and the Baryon asymmetry can be used to constrain the scale of 
Left-Right symmetry breaking.
\end{abstract}


\maketitle

\section{Beyond electroweak baryogenesis}
Explaining the observed baryon asymmetry of the Universe within the
framework of gauge theories and the standard Big Bang cosmology 
remains an open problem. The combination $B+L$ of the baryon and 
lepton numbers is known to be anomalous 
in the Standard Model (SM). 
For $T>T_{EW}$, the temperature of the electroweak phase 
transition, the $B+L$ violation becomes unspuppressed\cite{KRS,ArnMcl,aaps}.
Thus any $B+L$ generated at the scales of Grand Unified 
Theories (GUT's) would be erased. Further, all the known GUTs 
preserve $B-L$ whose natural value should be zero. Thus the GUT 
solution\cite{yoshimura,weinBgen} of baryogenesis is unlikely 
to be true.

Possible mechanisms for generating observed 
baryon asymmetry at electroweak scale are reviewed 
in\cite{cknrev,trodrev,yajbgen}.
The strategy is to assume a first order phase transition
to ensure an epoch of non-equilibrium evolution,
during which the $B$, $C$ and $CP$ violating effects must
take place, satisfying the Sakharov criteria\cite{sakh}. 
However with Higgs mass as large as  $115$ GeV, 
the phase transition in SM would be second order, making 
baryogenesis unfeasible in SM. 

Mechanisms in which the non-equilibrium evolution is due
to the presence of topological defects, viz., domain walls \cite{mencoo}, 
and cosmic strings \cite{sbdanduay1,branden1} have also been
considered. Ensuring first order 
phase transition requires fine tuning the couplings and
particle content of the model, while existence of defects relies 
only on topological features of the  vacuum manifolds. 
An appealing possibiity is provided by the intermediate scale
unification in the Left-Right symmetric model. Large Majorana masses
for the neutrinos permit Lepton number violation and the resulting
baryogenesis can be used to constrain the scale of the large mass 
from astrophysical data on the neutrino mass scale.

\section{Domain walls and L violation in Left-Right symmetric model}
The Left-Right symmetric model  consists of an 
additional group $SU(2)_R$, under
which the $e^{-}_R$ possesses a partner $\nu_R$. A new
hypercharge needed turns out to be $B-L$ and
the formula for the electric charge reads \cite{mohapbook,Senjanovic2}
\(
Q\ =\  T^3_L + T^3_R + {1\over 2}(B-L).
\)
The neutrinos naturally possess majorana masses permitting 
unrestricted $L$ violation if the symmetry breaking 
is achieved using two Higgses $\dr$, $\dl$, triplets under the
groups suggested by the respective subscripts. A bidoublet $\p$ 
is also needed and contains two copies of the SM type Higgs. 
The breakdown
$SU(2)_R\otimes U(1)_{B-L}$$\rightarrow U(1)_Y$ also signals
breaking of the discrete symmetry $SU(2)_L \leftrightarrow SU(2)_R$ 
which gives rise to domain walls\cite{ywmmc,lazar,lew-rio} separating two 
kinds of degenerate  vacuua. It also signals $C$ violation. 
$CP$ violation arises out of complex expectation values for the 
Higgs and finally the motion of the walls must be directional 
in order to produce a final state with $SU(2)_L$ unbroken.
Thus the Sakharov criteria are satisfied when the 
$L\leftrightarrow R$ violating phase transition occurs. 

\section{Leptogenesis and baryogenesis}
\begin{figure}

{\par\resizebox*{0.45\textwidth}{!}
{\rotatebox{-90}{\includegraphics{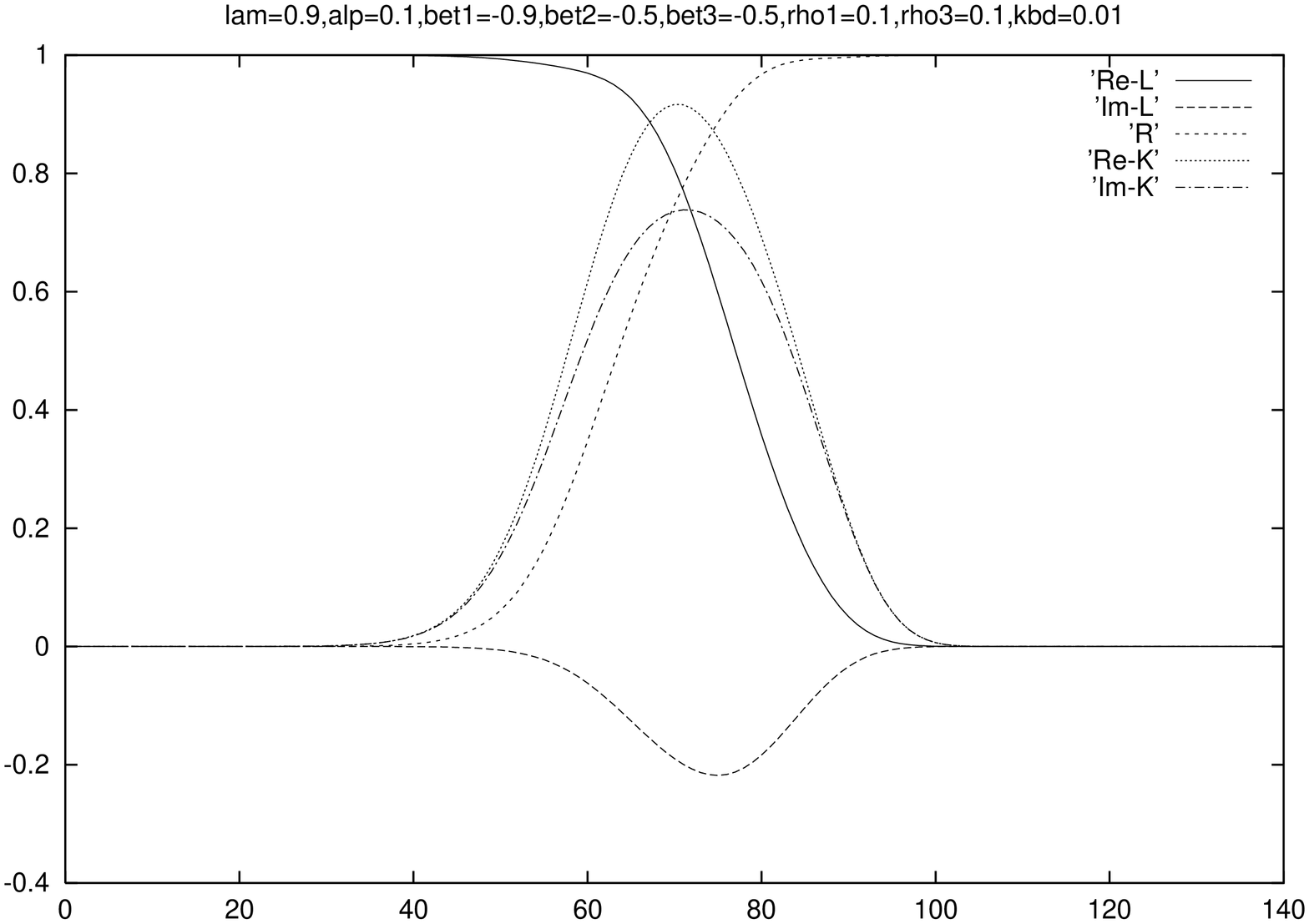}}}
\hfil
\resizebox*{0.45\textwidth}{!}
{\rotatebox{-90}{\includegraphics{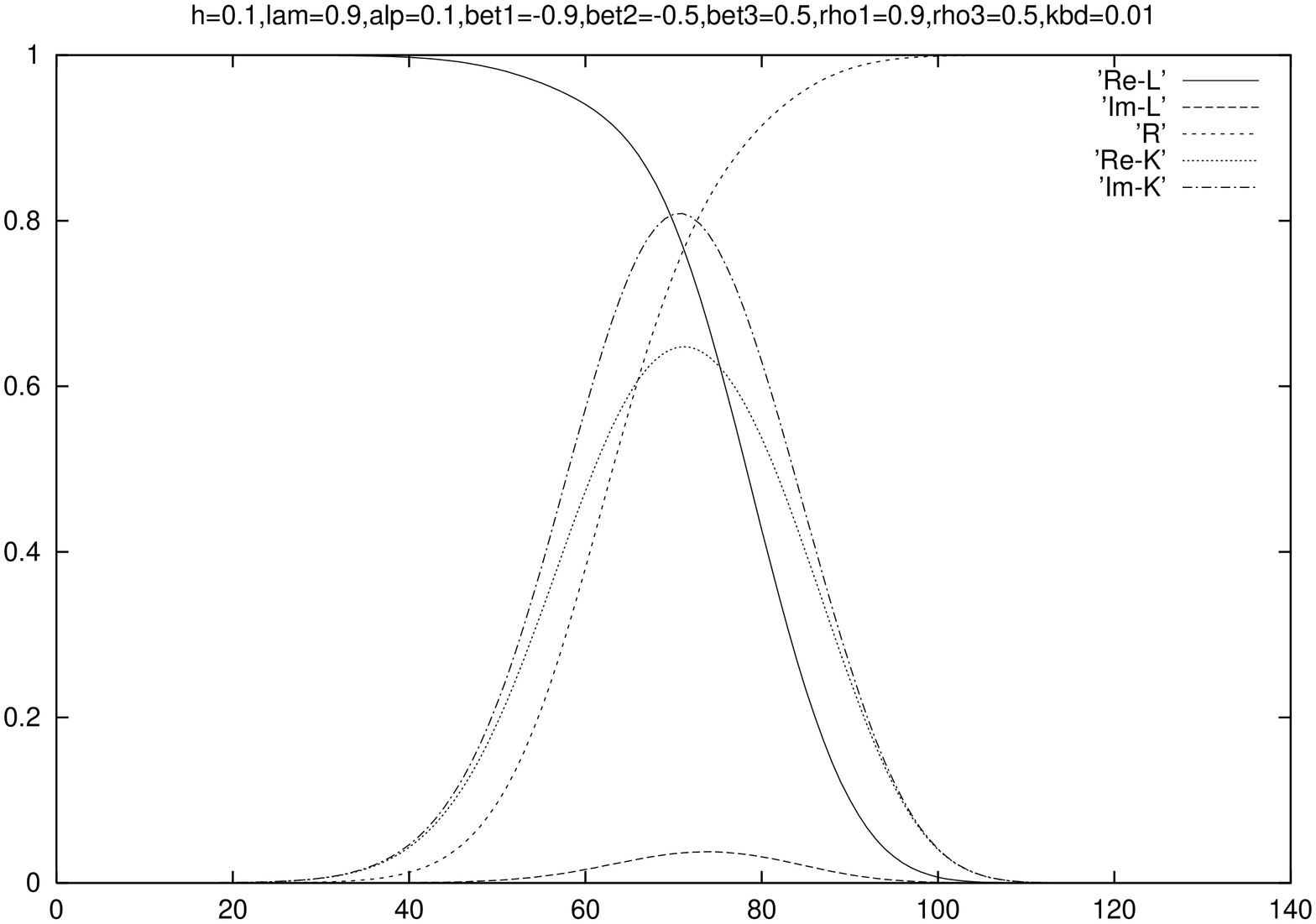}}}
\par}

\caption{Scalar field condensates for two sets of parameter values}
\label{fig:profiles}
\end{figure}

At least two of the Higgs expectation values in L-R model
are generically complex, thus providing natural $CP$
violation\cite{dgko} permitting all parameters in the Higgs 
potential to be real. Within the thickness  of the domain 
wall the $CP$ violating phase becomes position
dependent. Under these circumstances a formalism exists
\cite{jpt,clijokai,clikai}, wherein the chemical potential
$\mu_\Ls$ created for the Lepton number can be computed as 
a solution of the diffusion equation
\be
\label{eq:diffeq}
-D_\nu \mu_\Ls'' - v_w \mu_\Ls'
+ \theta(x)\, \Gamma_{\rm hf}\,\mu_\Ls = S(x).
\ee
Here $D_\nu$ is the neutrino diffusion coefficient,
$v_w$ is the velocity of the wall, taken to be moving in the $+x$
direction, $\Gamma_{\rm hf}$ is the rate of helicity
flipping interactions taking place in front of the wall (hence
the step function $\theta(x)$), and $S$ is the source term
which contains derivatives of the position dependent complex 
Dirac mass.
In \cite{cynr} the existence of such a position dependent phase was
established on general grounds and in a few numerical examples,
two of which are included in Fig. \ref{fig:profiles}.
Here the expectation values of the $\dl$, denoted $L$ and that
of the bidoublet denoted $K$ are taken to be complex while that
of $\dr$, denoted $R$ is real. 

After integration of the above equation and using inputs 
from the numerical solutions we find raw Lepton aymmetry
expressed as a ratio to the entropy density,
\be
\eta^{\rm raw}_{\sss L} \cong 0.01\,  v_w {1\over g_*}\, 
	{M_{\sss N}^4\over T^5\Delta_w}
\label{eq:ans2}
\ee
with $M_{\sss N}$ the majorana neutrino mass, $\Delta_w$ standing 
for the wall width and $g_*$ the effective thermodynamic degrees of
freedom at the epoch with temperature $T$.
This undergoes depletion due to $L$ violating processes which
are in equilibrium at that epoch. However the high temperature
sphalerons are efficiently converting the $L$ asymmetry into
$B$ asymmetry according to the chemical potential balance
in SM \cite{khlebshap}, given by
\be
	\Delta n_{B}\ =\ {28\over 79} \Delta n_{B-L}\ =\ 
-{28\over 51} \Delta n_L.
\ee
After these processes are all taken into account, we require
$B$ asymmetry $\eta_B$ to be $10^{-d_{\sss B}}$,  observed 
$d_{\sss B}$ being $\sim 11$,
leading to the following predictions. If the heaviest neutrino 
mass is $1$ eV, for example, the temperature of the
$LR$ phase transition is predicted to be 
\be
\label{limit1}
	T_{\sss LR}\  \lsim 10^{13} {\rm\ GeV} \times \left({{\rm\ eV}\over m_\nu }\right)^2
	\times \left({d_{\sss B}\over 10}\right)
\ee
If we are in the opposite regime, $M_\Ns < T_{\sss LR}$,
the bound on the heaviest neutrino mass is
\be
\label{limit2}
m_\nu < 0.3\ {\rm eV}\times \left({d_{\sss B}\over 10}\right)
\ee
It is interesting that this value is compatible with, and not very far from
the value implied by atmospheric neutrino observations.

Thus a hitherto unexplored mechanism exists in the Left-Right
symmetric model for generation of the observed baryon asymmetry of the
Universe. Further impliations and detailed discussion are contained in
\cite{cynr}.

\end{document}